\documentclass[10pt,a4paper]{article}

\usepackage[T1]{fontenc}     
\usepackage{times}           
\usepackage{microtype}       
\usepackage[left]{lineno}

\usepackage{amsmath,mathtools,amssymb}  
\usepackage{graphicx,relsize}           
\usepackage{multirow,booktabs,threeparttable} 
\usepackage{tabularx,ragged2e}          
\usepackage{siunitx}                    
\usepackage[dvips]{xy}                  

\usepackage{setspace}                   
\selectfont               
\usepackage{xcolor}                     
\usepackage{titlesec,etoolbox}          

\titleformat{\section}
  {\fontsize{10}{10}\selectfont\bfseries\raggedright} 
  {\thesection.}{0.3em}{}                              
\titlespacing*{\section}{0pt}{10pt}{6pt}               

\titleformat{\subsection}
  {\fontsize{10}{10}\selectfont\bfseries\raggedright}
  {\thesubsection.}{0.3em}{}
\titlespacing*{\subsection}{0pt}{10pt}{6pt}

\titleformat{\subsubsection}
  {\fontsize{10}{10}\selectfont\bfseries\itshape\raggedright}
  {\thesubsubsection.}{0.3em}{}
\titlespacing*{\subsubsection}{0pt}{10pt}{3pt}

\usepackage[numbers,sort&compress]{natbib}  
\usepackage[colorlinks=true,
            linkcolor=blue,
            citecolor=blue,
            urlcolor=magenta]{hyperref}     
\usepackage{url}
\usepackage{doi}
\usepackage[left=2cm, right=2cm, top=2cm, bottom=2cm]{geometry}

\title{Interspecific information use facilitates species coexistence in ecosystems}
\author{
    Wei Tao\textsuperscript{1$\dagger$}, 
	Ju Kang\textsuperscript{2,$\dagger$}, 
    Wenxiu Yang\textsuperscript{1}, 
    Yiyuan Niu\textsuperscript{1},
    Xin Wang\textsuperscript{1,*}
}
\date{}
\newcommand{\affil}[2]{%
	\textsuperscript{#1}#2%
}

\makeatletter
\renewcommand{\maketitle}{
	\begin{center}
		{\LARGE\bfseries \@title\par}
		\vspace{1em}  
		{\normalsize \@author\par}
		\vspace{-0.3em} 
		\setlength{\parskip}{1pt}
		\@date
	\end{center}
}
\makeatother
\begin{document}
	\maketitle
	
	\begin{center}
		\affil{1}{School of Physics, Sun Yat-sen University, Guangzhou 510275, China}\\
		\affil{2}{School of Ecology, Sun Yat-sen University, Shenzhen 518107, China}\\
		\affil{$\dagger$}{These authors contributed equally to this work}\\		
		\affil{*}{Corresponding author: \href{mailto:wangxin36@mail.sysu.edu.cn}{wangxin36@mail.sysu.edu.cn}}
	\end{center}	
	\begin{abstract}
		\noindent Explaining how competing species coexist remains a central question in ecology. The well-known competitive exclusion principle (CEP) states that two species competing for the same resource cannot stably coexist, and more generally, that the number of consumer species is bounded by the number of resource species at steady state. However, the remarkable species diversity observed in natural ecosystems, exemplified by the paradox of the plankton, challenges this principle. Here, we show that interspecific social information use among predators provides a mechanism that fundamentally relaxes the constraints of competitive exclusion. A model of predation dynamics that incorporates interspecific information use naturally explains coexistence beyond the limits imposed by CEP. Our model quantitatively reproduces two classical experiments that contradicts the CEP and captures coexistence patterns documented in natural ecosystems, offering a general mechanism for the maintenance of biodiversity in ecological communities.
		\vspace{0.3cm}			
	\end{abstract}	
\section{Introduction}

Understanding how multiple species coexist within shared environments remains a fundamental question in ecology~\cite{Pennisi2005}, as it underpins both the maintenance of biodiversity and the stability of ecosystems. The classical competitive exclusion principle (CEP)~\cite{Hardin1960,Gause1934}, however, posits that species occupying identical niches cannot stably coexist over the long term. In its simplest form, the CEP predicts that when two consumer species compete for a single limiting resource, one will inevitably exclude the other. More generally, consumer–resource theory extends this reasoning, asserting that the number of consumer species capable of coexisting at steady state cannot exceed the number of distinct resource 
types~\cite{MacArthur1964,Levin1970,McGehee1977}. Nevertheless, numerous observations from natural ecosystems challenge the CEP. In aquatic environments, for example, a vast diversity of plankton species coexist despite limited resources, a phenomenon known as the “paradox of the plankton”~\cite{deVargas2015,Hutchinson1961}. Likewise, tropical forests harbor thousands of tree species~\cite{Hoorn2010}; grassland systems sustain intricate plant and herbivore coexistence networks~\cite{Olff2002,McNaughton1985}; and even the resource-scarce Antarctic supports thriving communities of penguins and seabirds that depend on krill populations~\cite{Croxall2004}. These widespread examples reveal a fundamental mismatch between classical ecological theory and the observed richness of biodiversity in nature. Explaining how multiple species coexist despite apparent competitive constraints therefore remains one of the central challenges in ecology~\cite{Pennisi2005}.

Since the seminal mathematical formulation of the CEP by MacArthur and Levins~\cite{MacArthur1964} in the 1960s, ecologists have sought to understand how natural communities sustain the remarkable diversity observed in real ecosystems. Over the decades, a broad range of mechanisms has been proposed to relax the constraints imposed by the CEP. One line of research emphasizes that natural ecosystems often violate the steady-state assumptions required for the CEP to hold. Temporal fluctuations~\cite{Hutchinson1961,Levins1979,Bloxham2024}, spatial heterogeneity~\cite{Levin1974a,Gupta2021,VillaMartin2020}, and self-organized oscillatory or chaotic population 
dynamics~\cite{Levin1974b,Huisman1999} can prevent communities from reaching equilibrium, thereby enabling coexistence despite competition for limited resources. Another line of research focuses on biological interactions and evolutionary processes that actively promote coexistence, including predator interference~\cite{Beddington1975,DeAngelis1975,Kang2024a,Kang2024b}, toxins~\cite{Czaran2002}, coevolution~\cite{Xue2017}, metabolic trade-offs~\cite{Posfai2017,Weiner2019}, rock–paper-scissors~\cite{Kelsic2015,Kerr2002} cross-feeding~\cite{Goyal2018,Goldford2018,Niehaus2019}, pack hunting~\cite{Wang2020}, “kill-the-winner”~\cite{Thingstad2000}, and collective behavior~\cite{Dalziel2021}. These and other complex ecological interactions~\cite{Grilli2017,Ratzke2020,Kang2026,Ruxton1992} reveal diverse pathways to coexistence, yet our understanding remains incomplete.

While many mechanisms have been proposed to explain coexistence beyond the constraints of the CEP, one potentially important yet underexplored factor is the use of social information by predators. Social information, acquired from conspecifics or heterospecifics at the same trophic level, can enhance foraging efficiency and is a widespread phenomenon across taxa~\cite{Goodale2010,Hamalainen2023,Seppanen2007,Dall2005}. For example, seabirds such as \textit{Kittiwakes} and \textit{Albatrosses} emit informative cues that guide other seabird species toward productive foraging grounds~\cite{Seppanen2007,Monier2024,Silverman2001,Silverman2004,Hoffman1981}; \textit{Sticklebacks} use information on the foraging success of heterospecific fishes to identify food-rich areas~\cite{Seppanen2007,Coolen2003}; \textit{Stingless bees} follow other bee species to locate floral resources~\cite{Seppanen2007,Gloag2021}; and bats extract information from the echolocation calls of heterospecific bats to optimize their foraging strategies~\cite{Culina2019,Lewanzik2019}. Despite extensive studies on social information use, its potential role in mitigating competitive exclusion and promoting coexistence remains largely unexplored.

In this study, motivated by principles of chemical reaction kinetics~\cite{Ruxton1992,Huisman1997,Wang2020} and grounded in MacArthur's consumer--resource framework~\cite{MacArthur1969,MacArthur1970,Chesson1990}, we develop a model that extends our previously established chasing-pair formulation~\cite{Wang2020} to incorporate interspecific information use. The chasing-pair model describes pairwise encounters between consumers and resources and, by itself, remains constrained by the CEP. Incorporating information use fundamentally alters this outcome: two consumer species can stably coexist while exploiting a single resource. The resulting coexistence state is robust to stochastic fluctuations and therefore may emerge in natural ecosystems. Notably, our model naturally explains two classical experiments that contradict the CEP~\cite{Park1954,Ayala1969} and quantitatively accounts for species coexistence patterns observed in natural ecological communities~\cite{Hatch1988,Hatch1989}.

\section{Results}
\subsection{A model incorporating interspecific information use}
Here, we develop a model based on MacArthur’s consumer–resource framework~\cite{MacArthur1969,MacArthur1970,Chesson1990}, which integrates interspecific social information use into the chasing-pair scenario~\cite{Wang2020}, explicitly accounting for pairwise encounters between individual consumers and resources. In the simplest case, two consumer species, $C_1$ and $C_2$, compete for a single resource species $R$. Consumers are biotic, whereas the resource may be either biotic or abiotic. All individuals move randomly in space until a consumer and a resource come into proximity, whereupon the consumer initiates a chase, forming a chasing pair, denoted as $C_i^\text{(P)}\vee R^{\text{(P)}}$, where the superscript “(P)” indicates the paired state. As in our previous studies~\cite{Kang2024a,Kang2024b,Wang2020}, a system involving only chasing pairs (Fig.~\ref{Abiotic_resources}A) can be described as follows:			
\begin{equation}
\begin{cases}
\dot{x}_i = a_i C_i^\text{(F)} R^\text{(F)} - (k_i + d_i)x_i, \\[6pt]
\dot{C}_i = w_i k_i x_i - D_i C_i, \\[6pt]
\dot{R} = g\big(\{R\}, \{x_i\}, \{C_i\}\big) & (i = 1,2).
\end{cases}
\label{eq1}
\end{equation}
Here,  $x_i$ represents the chasing pair, while symbols with the superscript “(F)” denote freely wandering populations ($i=1, 2$). The parameters $a_i$, $d_i$ and $k_i$ represent the encounter rate, escape rate, and capture rate of the predation process, respectively. $w_i$ is the biomass conversion ratio from species  $R$ to $C_i$, and $D_i$ is the mortality rate of $C_i$. The function $g(\{R\}, \{x_i\}, \{C_i\})$ remains unspecified. The total population abundances of the consumer species $C_i$ and the resource species $R$ are given by $C_i=C_i^\text{(F)}+x_i$ and $R=R^\text{(F)}+x_1+x_2$, respectively.

Motivated by the observation that a wide range of predator species such as birds, fish, bats, and insects can exploit social information from heterospecifics to enhance their searching efficiency~\cite{Goodale2010,Seppanen2007,Dall2005,Monier2024,Silverman2001,Silverman2004,Hoffman1981,Coolen2003,Gloag2021,Culina2019,Lewanzik2019}, we explicitly incorporate the effect of interspecific social information use into our model. For the case in which consumer species $C_1$ can use information from $C_2$ to improve its search efficiency, but not vice versa (Fig.~\ref{Abiotic_resources}B), the model retains the same structure as Eq.~\ref{eq1}, except that the encounter rate $a_1$ becomes a function of $C_2$ (while $a_2$ remains constant):
\begin{equation}
a_1(C_2) = a'_1 \left[ 1 + \dfrac{l_2 C_2}{K_2 + C_2}\right].
\label{eq2}
\end{equation}
We assume a Monod type formulation ~\cite{Monod1949} to describe the enhancement of search efficiency through social information use. Here, $l_2$ denotes the maximum relative increase in searching efficiency, $a'_1$ is the encounter rate in the absence of social information transfer, and $K_2$ represents the half saturation constant at which this increase levels off.

To proceed with the formulation, we specify the population dynamics of the resource following the same structural principles as those in MacArthur’s framework ~\cite{MacArthur1964,MacArthur1970,Chesson1990}. Accordingly, the function $g(\{R\}, \{x_i\}, \{C_i\})$ is expressed as

\begin{equation}
g\big(\{R\}, \{x_i\}, \{C_i\}\big) =
\begin{cases}
\eta R\left(1 - \dfrac{R}{\kappa_0}\right) - k_1 x_1 - k_2 x_2 & (\text{for biotic resources}); \\[8pt]
\zeta \left(1 - \dfrac{R}{\kappa_a}\right) - k_1 x_1 - k_2 x_2 & (\text{for abiotic resources}).
\end{cases}
\label{eq3}
\end{equation}	
In the absence of consumers, the population of a biotic resource follows logistic growth, characterized by an intrinsic rate $\eta$ and a carrying capacity $\kappa_0$. In contrast, an abiotic resource varies according to an external supply rate $\zeta$, with $\kappa_a$ denoting the equilibrium abundance maintained in the absence of consumers.

In our model incorporating the effect of interspecific information use, the population dynamics follow Eqs.~\ref{eq1}- \ref{eq3}. To facilitate analysis, we used dimensional analysis to express all parameters in dimensionless form (see SM sec.IV). For clarity, the same symbols are used throughout, and all parameters are treated as dimensionless unless stated otherwise.
	
\subsection{Interspecific information use facilitates the breakdown of the CEP}

\begin{figure}[ht!]
	\centering
	\includegraphics[width=12cm]{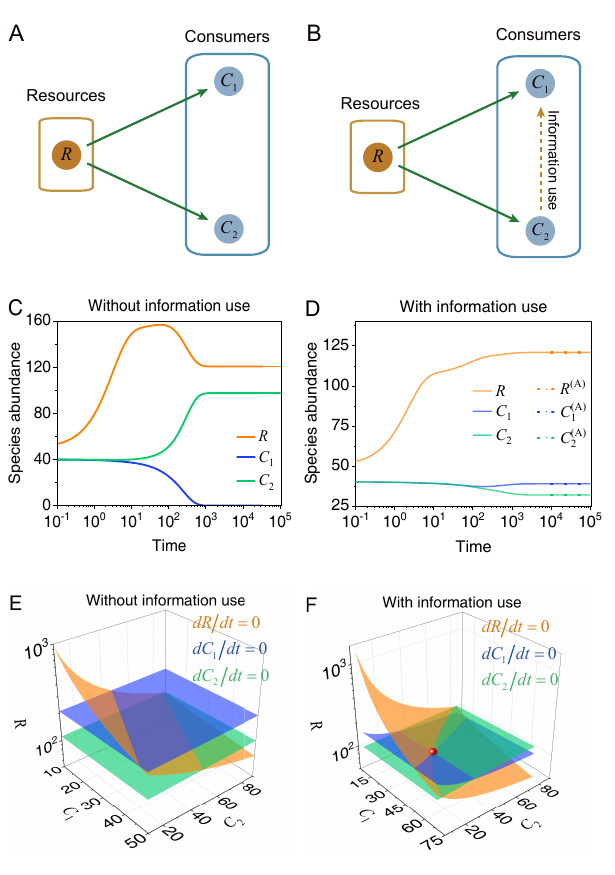}
	\caption{\label{Abiotic_resources}
         Interspecific information use facilitates coexistence beyond the CEP limit. (A) Schematic illustration of two consumer species competing for a single resource. Green arrows indicate the energy (biomass) flow within the trophic interaction.(B) Model incorporating interspecific information use based on the schematic in (A), where consumer species $C_1$ use social information from $C_2$ to enhance its resource-searching efficiency.(C-D) Temporal dynamics of two consumer species competing for a single abiotic resource. (E-F) Positive steady-state solutions derived from the equilibrium equations (see Eqs.~\ref{eq1}-\ref{eq3}): $\dot{R}=0$ (orange surface),  $\dot{C_1}=0$ (blue surface), and $\dot{C_2}=0$ (green surface), representing the zero-growth isoclines. The red dot denotes the stable fixed point. The dotted curves in (D) indicate the analytical steady-state abundances (denoted by superscript '(A)'). See SM Sec. VI and Table S1 for simulation details of Figs. 1–3.}
    \end{figure}

In our previous studies~\cite{Kang2024a,Wang2020}, we showed that the scenario involving only chasing pairs is constrained by the CEP. Consistent with this finding, as shown in Figs.~\ref{Abiotic_resources}C and S1A, when interspecific information transfer is absent, two consumer species fail to coexist at constant population densities while competing for a single resource type. The zero-growth isolines of the consumer species ($\dot{C_l}=0$) satisfy $f_i (R^{(F)})=D_i$  $(i=1,2)$~\cite{Kang2024a,Wang2020}(see SM Sec. III for details), forming two parallel surfaces in the ($C_1,C_2,R$) space (blue and green surfaces in Figs.~\ref{Abiotic_resources}E and S1C), which prevents stable coexistence. 

\begin{figure}[ht!]
	\centering
	\includegraphics[width=14cm]{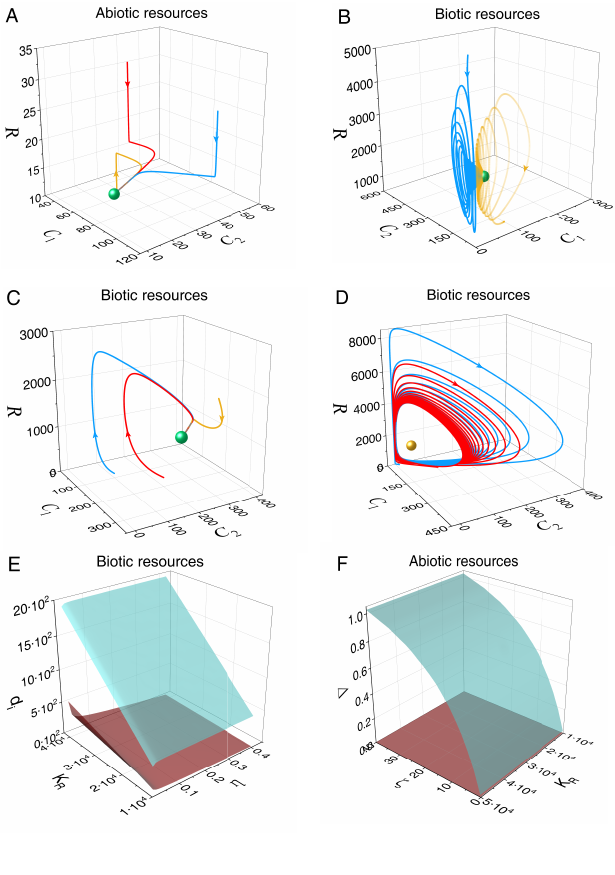}
	\caption{\label{Chaos}
        Coexistence modes and corresponding parameter regions. (A-D) Representative coexistence trajectories in state space. (A) Abiotic resource case: the coexistence equilibrium (green dot) serves as a global attractor. (B-D) Biotic resource cases: in (B) and (C), the coexistence equilibria (green dots) remain globally stable, whereas in (D) the equilibrium (yellow dot) loses stability, and trajectories converge to a stable limit cycle. (E-F) Parameter regions supporting coexistence in the ODE studies. The parameter space bounded by the blue and red surfaces corresponds to stable coexistence. $\Delta$ is defined as $\Delta  = {{\left( {{D_1} - {D_2}} \right)} \mathord{\left/{\vphantom {{\left( {{D_1} - {D_2}} \right)} {{D_2}}}} \right. \kern-\nulldelimiterspace} {{D_2}}}$ in (F).} 	
    \end{figure}

In contrast, incorporating interspecific information use alters the geometry of the phase-plane diagram. The zero-growth surfaces of the consumers and the resource transform into three non-parallel planes that converge at a common point (Figs.~\ref{Abiotic_resources}F and S1D).  This intersection forms a stable fixed point, allowing two consumer species to coexist stably with a single resource (Figs.~\ref{Abiotic_resources}D and S1B) and thereby breaking the constraint imposed by the CEP. This mechanism supports multiple coexistence modes. For biotic resources, the coexistence state may be globally attractive with a stable fixed point (Fig.~\ref{Chaos}B and \ref{Chaos}C). Alternatively, an unstable intersection can drive oscillatory coexistence, giving rise to a limit cycle in phase space (Fig.~\ref{Chaos}D). For abiotic resources, only the stable fixed point persists and the trajectories in phase space remain globally attractive. Notably, the analytically derived stable coexistence states (see Eqs.S3-S5 and S9) match the simulation results perfectly (Figs.~\ref{Abiotic_resources}D and S1B).
	\begin{figure}[!htbp]
	\centering
	\includegraphics[width=14.5cm]{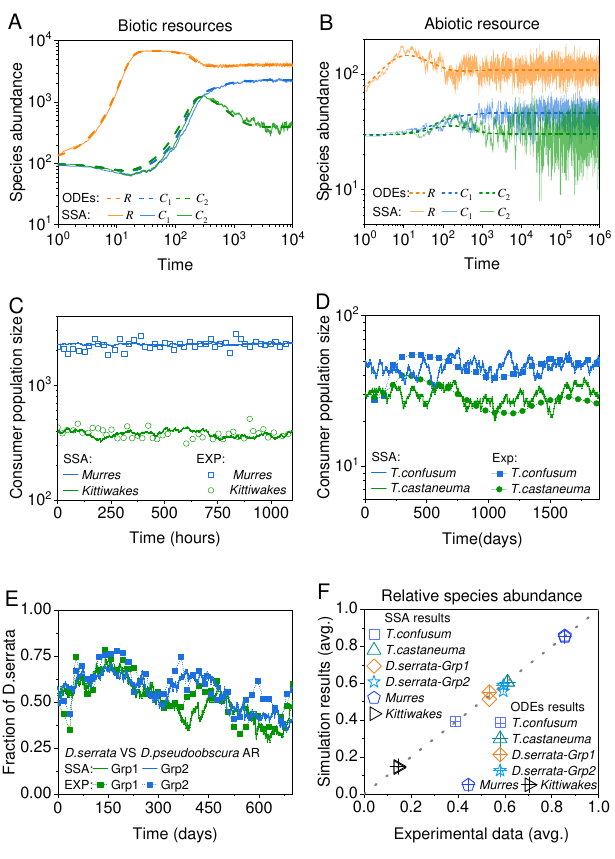}
	\caption{\label{Experiment} 
		Interspecific information use explains experimental observations that contradict the CEP. (A-B) Representative temporal dynamics simulated using ODEs and SSA.  
		(C) Comparison of SSA results with field survey data shows that two seabird species, \textit{Black legged Kittiwakes} and \textit{Murres}, coexist on the same cliff surfaces~\cite{Hatch1988,Hatch1989}. 
		Information used by \textit{Black legged Kittiwakes} facilitates prey detection by \textit{Murres}, helping to explain this phenomenon.  
		(D-E) SSA simulations reproduce two classical experiments that violate the CEP. Two \textit{Tribolium} species or two \textit{Drosophila} species coexist with a single resource type in each experiment~\cite{Park1954,Ayala1969}.
        In (C–E), the relative root mean square errors (rRMSEs) between SSA simulations and experimental data are as follows: \textit{Murres}, 0.09; \textit{Black legged Kittiwakes}, 0.13; \textit{T. confusum}, 0.18; \textit{T. castaneum}, 0.18; \textit{D. serrata} Grp1, 0.15; and \textit{D. serrata} Grp2, 0.15.
        (F) Comparison of time averaged relative population abundances between SSA results and experimental data~\cite{Park1954,Hatch1988,Hatch1989,Ayala1969}. All data points lie close to the dashed line, where theoretical results equal experimental observations. The Pearson correlation coefficients between experimental data and SSA and ODE results are 0.998 and 0.997, respectively. See SM Sec.V and VI for details of the SSA simulations.} 	
\end{figure}%

To confirm that this facilitated coexistence does not arise from accidental parameter choices, we systematically explored the parameter space allowing stable coexistence between two consumer species and the resource. As shown in Fig.~\ref{Chaos}E and \ref{Chaos}F, for both biotic and abiotic cases, the region below the blue surface and above the red surface represents stable steady coexistence. In both scenarios, a finite parameter region allows the breakdown of the CEP, demonstrating that this relaxation of the CEP constraint is a robust property of the system rather than a consequence of a pathological parameter set.
\subsection{Robustness of the facilitated coexistence state to stochasticity.}	
Stochasticity is an inherent feature of natural ecosystems, yet it frequently undermines species coexistence~\cite{Xue2017}. Classical mechanisms such as “kill the winner” lose effectiveness once stochasticity is taken into account~\cite{Xue2017}. To assess the influence of stochasticity on our model, we employed the stochastic simulation algorithm (SSA)~\cite{Gillespie2007}. Remarkably, as shown in Fig.~\ref{Experiment}A-B, the coexistence state facilitated by interspecific information use remains resilient to stochastic perturbations, with both consumer species persisting over time in the SSA. These findings demonstrate that interspecific information use robustly sustains species coexistence even under stochastic conditions.
\subsection{ Intuitive understanding: information source facilitates coexistence}
Intuitively, the relaxation of the competitive exclusion principle through interspecific information use can be understood from an information resource perspective. In the simplified case shown in Fig.~\ref{Abiotic_resources}B, consumer species $C_1$ exploits the social information provided by species $C_2$ to improve its search efficiency for the shared resource species $R$.Although $R$ remains the sole biomass source consumed by both consumer species, $C_2$ simultaneously acts as an information source, effectively functioning as an additional type of resource. This dual role facilitates species coexistence in a manner analogous to toxin-mediated coexistence.

From a quantitative perspective, this facilitation can be interpreted through the steady-state constraints governing population dynamics. Without information use, the steady-state condition for the consumer species in the chasing-pair scenario requires $f_i(R^\text{(F)})/D_i=1$ $(i=1,2)$
 (see Eq.S14 in SM Sec. III and Ref.~\cite{Wang2020} for details). Geometrically, this condition corresponds to two parallel surfaces(the blue and green surfaces in Fig.~\ref{Abiotic_resources}E) in the $(C_1,C_2,R)$ coordinate space, which do not intersect at a common point and therefore do not allow a feasible coexistence equilibrium. However, when interspecific information use is introduced, the steady-state requirement for $C_1$ becomes $F_1(R^\text{(F)},C_2)/D_1=1$ (see Eq. S15 in SM Sec. III for details). Here, the presence of $C_2$ as an information source introduces an additional degree of freedom into the constraint. Consequently, the steady-state conditions of the three species correspond to three non-parallel surfaces that can intersect at a common point (Fig.~\ref{Abiotic_resources}F), thereby enabling stable species coexistence.

\subsection{Quantitative explanation of observations across diverse ecological systems}

By incorporating interspecific information use into the consumer–resource framework, our model quantitatively reproduces coexistence patterns across diverse ecosystems, circumventing the constraint of the CEP. It captures the stable coexistence of seabird species such as \textit{Murres} and \textit{Black legged Kittiwakes}~\cite{Hatch1988,Hatch1989}(Fig.~\ref{Experiment}C), where behavioral cues about prey abundance facilitate efficient resource sharing~\cite{Hoffman1981}. The model also accounts for two classical laboratory experiments that contradict the CEP: the long-term coexistence of \textit{Tribolium castaneum} and \textit{T. confusum}~\cite{Ayala1969}(Fig.~\ref{Experiment}D), possibly enabled by cross-species detection of aggregation pheromones such as 4,8-dimethyldecanal~\cite{Kim2005,Suzuki1980}, and the persistence of mixed \textit{Drosophila} communities~\cite{Park1954} (Fig.~\ref{Experiment}E), potentially mediated by behavioral information use~\cite{Girardeau2021}. Model predictions closely align with empirical data (Fig.~\ref{Experiment}C–E), with residual fluctuations primarily reflecting environmental stochasticity. Both deterministic (ODE) and stochastic (SSA) simulations reproduce observed coexistence patterns across systems (Fig.~\ref{Experiment}F), highlighting the predictive power and generality of this mechanism in structuring ecological communities.
\section{Discussion}

The contradiction between the CEP and the high species diversity observed in nature has long been a central puzzle in ecology~\cite{Pennisi2005}. Although various mechanisms have been proposed to relax the constraints imposed by the CEP~\cite{Hutchinson1961,Levins1979,Levin1974a,Levin1974b,Huisman1999,Beddington1975,DeAngelis1975,Kang2024a,Czaran2002,Xue2017,Posfai2017,Weiner2019,Kelsic2015,Kerr2002,Goyal2018,Goldford2018,Niehaus2019,Wang2020,Dalziel2021,Grilli2017,Ratzke2020,Kang2026,Thingstad2000}, our understanding of how species coexistence is maintained remains incomplete. Here we identify an overlooked factor, interspecific social information use, as a mechanism capable of relaxing the limits of the CEP. By explicitly incorporating information use into a consumer-resource framework, our analytical results and simulations based on both ODEs and SSA demonstrate that two consumer species can stably coexist while exploiting a single resource. The coexistence state may take the form of either a stable fixed point or a stable oscillatory limit cycle, and it persists across broad parameter ranges. Its robustness to parameter variation and stochastic fluctuations suggests that the coexistence mechanism uncovered here is not a mathematical artifact but a biologically plausible phenomenon that can operate under realistic environmental conditions. Furthermore, our model quantitatively reproduces both classical laboratory experiments that violate the CEP~\cite{Park1954,Ayala1969} and field observations of coexistence among wild populations~\cite{Hatch1988,Hatch1989}.

Interspecific information use is widespread across taxa~\cite{Goodale2010,Seppanen2007,Dall2005,Hamalainen2023}. It has been documented in diverse organisms, including birds~\cite{Seppanen2007,Monier2024,Silverman2001,Silverman2004,Hoffman1981,Forsman2022,Forsman2011,Seppanen2011}, fish~\cite{Seppanen2007,Coolen2003}, bats~\cite{Culina2019,Lewanzik2019}, and insects~\cite{Seppanen2007,Gloag2021,Loukola2020}, and is increasingly recognized as an important component of ecological and evolutionary dynamics~\cite{Gil2019,Kane2014,Hein2020,Ashby2022,Danchin2011}. Previous theoretical studies have incorporated information use within generalized Lotka-Volterra frameworks~\cite{Gil2019} and omnivory networks~\cite{Kane2014}. The former implicitly assumes that the number of consumer species cannot exceed that of resource species (see ref.~\cite{Wang2020} for mathematical proof), while the latter explicitly imposes this constraint by construction. As a result, both approaches remain bound by the CEP and cannot explain coexistence among multiple consumers sharing a single resource. In contrast, by explicitly modeling the predation process and the behavioral coupling introduced by information use, our framework demonstrates how interspecific information sharing can effectively break the classical CEP constraint.

The implications of this mechanism may extend beyond observable ecosystems. In microbial communities, analogous forms of interspecific information use can arise through quorum sensing, where microbes communicate via diffusible chemical signals to coordinate gene expression~\cite{Miller2001}. Such signaling could, in principle, mediate interspecific coordination in resource exploitation and defense strategies. However, caution is warranted in extending our model directly to microbial systems, as microbial coexistence often involves additional processes such as cross feeding~\cite{Goyal2018,Goldford2018,Niehaus2019}, overflow metabolism~\cite{Wang2025}, and evolutionarily shaped substrate preferences~\cite{Wang2019}. These metabolic and evolutionary interactions complicate the identification of dominant coexistence mechanisms. Integrating information mediated interactions with these processes offers a promising direction for understanding how diverse species coexist in complex ecological systems.

\section*{Acknowledgments}
This work was supported by the National Natural Science Foundation of China (No.12474207). 

\section*{Conflict of interest}
The authors declare the following financial interests/personal relationships which may be considered as potential competing interests: Xin Wang reports financial support was provided by the National Natural Science Foundation of China. If there are other authors, they declare that they have no known competing financial interests or personal relationships that could have appeared to influence the work reported in this paper.

\section*{Author contributions}
X.W. conceived the project and designed the study. All authors contributed to model development, theoretical analysis, numerical simulations, data analysis, and manuscript preparation. 

\section*{Data and code availability}
All data and code used in this study are available from the corresponding author upon reasonable request.
\bibliographystyle{unsrtnat}
\bibliography{references}

\newpage
\noindent\textbf{\LARGE{Supplementary Materials}}

\renewcommand \thesection {\Roman{section}}
\renewcommand \thesubsection {\Alph{subsection}}
\renewcommand \thesubsubsection {\arabic{subsubsection}}

\makeatletter
\renewcommand{\thefigure}{S\@arabic\c@figure}
\renewcommand{\theequation}{S\@arabic\c@equation}
\makeatother

\setcounter{section}{0}    
\setcounter{figure}{0}    
\setcounter{equation}{0}
\renewcommand\theequation{S\arabic{equation}}		






\section{Stability analysis of the fixed-point solution}
\label{Stability analysis}

The steady states satisfying \(\dot{x}_i = 0,\ \dot{C}_i = 0\ (i = 1, 2),\ \text{and}\ \dot{R} = 0\)
 are defined as  fixed points, denoted as \(E(x_i, C_i, R)\). Local stability was assessed using linear stability analysis, where a fixed point is deemed stable only if all eigenvalues of its Jacobian matrix (denoted as \(\lambda_i,\, i = 1, \ldots, 5\)) possess negative real parts.

To identify parameter domains permitting coexistence, we allowed \( D_i \ (i = 1, 2) \) to vary as the main difference between the two consumer species 
\( C_1 \) and \( C_2 \), apart from the information use term \(\left(\frac{l_2 C_2}{K_2 + C_2} \right)\). 
Consequently, \( D_i \) characterizes the degree of competitive asymmetry between the two consumers. 
As illustrated in Fig. 2E-F, the parameter space enclosed by the blue and red surfaces represents a region of stable coexistence. This finding indicates the presence of a finite nonzero domain in which species can persist together over time, thus challenging the CEP.
\section {Analytical solutions of the species abundances at steady state}	
\label{sec:Analytical solutions}
At steady state, since 
\(\dot{x}_i = 0, \  \ \dot{C}_i = 0, \ \text{and} \ \dot{R} = 0\), then,
\begin{equation}
\left\{
\begin{aligned}
a&_1' \left[ 1 + l_2 \frac{C_2}{K_2 + C_2} \right] C_1^\text{(F)} R^\text{(F)} = (k_1 + d_1) x_1, \\
a&_2 C_2^\text{(F)} R^\text{(F)} = (k_2 + d_2) x_2, \\
w&_1 k_1 x_1 = D_1 C_1, \\
w&_2 k_2 x_2 = D_2 C_2, \\
g&\left( \{R\}, \{x_i\} \{C_i\} \right) = 0\  (i=1,2),
\end{aligned}
\right.
\tag{S1}
\label{eq:S1}
\end{equation}
where \(C_i^\text{(F)} = C_i - x_i\), and \(R^\text{(F)} = R - x_1 - x_2\).  
Focusing on the second and fourth sub-equations of Eq.~\ref{eq:S1} and eliminating \(x_1\) and \(x_2\), we obtain:

\begin{equation}
R^\text{(F)} = \frac{(k_2 + d_2) D_2}{a_2 \left( w_2 k_2 - D_2 \right)}.
\tag{S2}
\label{eq:S2}
\end{equation}
By combining with the first and third sub-equations of Eq. \ref{eq:S1}, Eq. \ref{eq:S2}, and
\(
C_{1}^\text{(F)} = C_{1} - x_{1}
\)
, we can obtain:

\begin{equation}
C_{2} =
\frac{
K_{2} \left( w_{1}k_{1} - D_{1} \right) a_{1}'\frac{(k_2 + d_2) D_2}{a_2 \left( w_2 k_2 - D_2 \right)}  
- (k_{1} + d_{1}) D_{1} K_{2}
}{
(k_{1} + d_{1}) D_{1}
- (1 + l_{2}) (w_{1}k_{1} - D_{1}) a_{1}' \frac{(k_2 + d_2) D_2}{a_2 \left( w_2 k_2 - D_2 \right)}
}.
\tag{S3}
\label{eq:S3}
\end{equation}
Since \( g\left( \{R\}, \{x_1\}, \{x_2\} \right) \) has two forms corresponding to biotic resources and abiotic resources, 
we first discuss the simpler case corresponding to abiotic resources, namely 
\(\zeta (1 - R / K_R) - k_1 x_1 - k_2 x_2 = 0\).  
By combining the third and fourth sub-equations of Eq.~\ref{eq:S1}, we obtain:

\begin{equation}
C_1 = \frac{w_1}{D_1} 
\left[ 
\zeta \left( 1 - \frac{R}{K_R} \right) 
- \frac{D_2 C_2}{w_2} 
\right].
\tag{S4}
\label{eq:S4}
\end{equation}
By combining with \(\zeta (1 - R / K_R) - k_1 x_1 - k_2 x_2 = 0\), 
\(R^\text{(F)} = R - x_1 - x_2\), Eq.~\ref{eq:S2}, and the third and fourth sub-equations of Eq.~\ref{eq:S1}, we have:
\begin{equation}
R = 
\frac{
\frac{(k_2 + d_2) D_2}{a_2 \left( w_2 k_2 - D_2 \right)} + \frac{\zeta}{k_1}-
\left(
\frac{D_2}{k_1 w_2} - \frac{D_2}{w_2 k_2}
\right) C_2
}{
1 + \frac{\zeta}{k_1 K_R}
}.
\tag{S5}
\label{eq:S5}
\end{equation}
Then, we focus on the form corresponding to biotic resources, denoted as 
\(\eta R (1 - R / K_R) - k_1 x_1 - k_2 x_2 = 0\). 
Similarly, by combining the third and fourth sub-equations of Eq.~\ref{eq:S1}, we obtain:
\begin{equation}
C_1 = 
\frac{w_1}{D_1}
\left[
\eta R \left( 1 - \frac{R}{K_R} \right)
- \frac{D_2 C_2}{w_2}
\right].
\tag{S6}
\label{eq:S6}
\end{equation}
Similarly, by combining 
\(\eta R (1 - R / K_R) - k_1 x_1 - k_2 x_2 = 0\), 
\(R^\text{(F)} = R - x_1 - x_2\), Eq.~\ref{eq:S2}, and the third and fourth sub-equations of Eq.~\ref{eq:S1}, 
we obtain:
\begin{equation}
0 =
\frac{\eta}{k_1 K_R} R^2
+
\left( 1 - \frac{\eta}{k_1} \right) R
+
\frac{D_2 C_2}{w_2}
\left( \frac{1}{k_1} - \frac{1}{k_2} \right)
- \frac{(k_2 + d_2) D_2}{a_2 \left( w_2 k_2 - D_2 \right)}.
\tag{S7}
\label{eq:S7}
\end{equation}

By solving Eq.~\ref{eq:S7} for \(R\), we obtain:
\begin{equation}
R =
\frac{
- \left( 1 - \frac{\eta}{k_1} \right)
\pm
\sqrt{
\left( 1 - \frac{\eta}{k_1} \right)^2
- 4 \frac{\eta}{k_1 K_R}
\left[
\frac{D_2 C_2}{w_2}
\left( \frac{1}{k_1} - \frac{1}{k_2} \right)
- \frac{(k_2 + d_2) D_2}{a_2 \left( w_2 k_2 - D_2 \right)}
\right]
}
}{
2 \frac{\eta}{k_1 K_R}
}.
\tag{S8}
\label{eq:S8}
\end{equation}

Normally \(\eta / k_1 < 1\), and since \(R > 0\), we have:
\begin{equation}
R =
\frac{
- \left( 1 - \frac{\eta}{k_1} \right)
+
\sqrt{
\left( 1 - \frac{\eta}{k_1} \right)^2
- 4 \frac{\eta}{k_1 K_R}
\left[
\frac{D_2 C_2}{w_2}
\left( \frac{1}{k_1} - \frac{1}{k_2} \right)
- \frac{(k_2 + d_2) D_2}{a_2 \left( w_2 k_2 - D_2 \right)}
\right]
}
}{
2 \frac{\eta}{k_1 K_R}
}.
\tag{S9}
\label{eq:S9}
\end{equation}

\section {Intuitive understanding of why interspecies information use can break CEP}
\label{Intuitive understanding}
This study provides an intuitive explanation of how information use can promote biodiversity by considering two consumer populations competing for the single biotic or abiotic resource. The population dynamics of the system are described by Eq. 1, assuming that the consumption process is in fast equilibrium (i.e.\({\dot x_i}\)=0). Accordingly, a set of equations can be derived to determine \({x_i}\) based on the population sizes of each species:
\begin{equation}
\left\{
\begin{aligned}
a&_1' \left[ 1 + l_2 \frac{C_2}{K_2 + C_2} \right] C_1^\text{(F)} R^\text{(F)} = (k_1 + d_1) x_1, \\
a&_2 C_2^\text{(F)} R^\text{(F)} = (k_2 + d_2) x_2, \\
C&_1^\text{(F)} + x_1 = C_1, \\
C&_2^\text{(F)} + x_2 = C_2, \\
R&^\text{(F)} + x_1 + x_2 = R.
\end{aligned}
\right.
\tag{S10}
\label{eq:S10}
\end{equation}

To express \( C_i^\text{(F)} \), \( x_i \), and \( R^\text{(F)} \ (i = 1, 2)\) in terms of \( R^\text{(F)} \) and \( C_i\), and by combining Eq.~\ref{eq:S10}, we have:
\begin{equation}
\left\{
\begin{aligned}
C&_1^\text{(F)}=\frac{C_1}{1+\frac{a_1'[1+l_2C_2/(C_2+K_2)]}{k_1+d_1}R^\text{(F)}},\\
C&_2^\text{(F)}=\frac{C_2}{1+\frac{a_2}{k_2+d_2}R^\text{(F)}}, \\
x&_1=\frac{\frac{a_1'[1+l_2C_2/(C_2+K_2)]}{k_1+d_1}R^\text{(F)}}{1+\frac{a_1'[1+l_2C_2/(C_2+K_2)]}{k_1+d_1}R^\text{(F)}}C_1, \\
x&_2=\frac{\frac{a_2}{k_2+d_2}R^\text{(F)}}{1+\frac{a_2}{k_2+d_2}R^\text{(F)}}C_2, \\
R&=R^\text{(F)}+x_1+x_2.
\end{aligned}
\right.
\tag{S11}
\label{eq:S11}
\end{equation}

From Eq.~\ref{eq:S11}, it can be seen that $R^\text{(F)}$, $C_1$, and $C_2$ are a set of independent variables.
By definition, the functional responses for \(C_1\) and\(C_2\) are given by $F_1(R^\text{(F)},C_2) \equiv w_1 k_1 x_1 / C_1$ and $F_2(R^\text{(F)}) \equiv w_2 k_2 x_2 / C_2$, respectively.In the first four sub-equations of Eq.~\ref{eq:S10}, by eliminating $C_1^\text{(F)}$ and $C_2^\text{(F)}$, we have:
\begin{equation}
\left\{
\begin{aligned}
F&_1(R^\text{(F)},C_2)=\frac{w_1k_1a_1'[1+l_2C_2/(K_2+C_2)]R^\text{(F)}}{k_1+d_1+a_1'[1+l_2C_2/(K_2+C_2)]R^\text{(F)}},\\
F&_2(R^\text{(F)})=\frac{w_2k_2a_2R^\text{(F)}}{k_2+d_2+a_2R^\text{(F)}}.
\end{aligned}
\right.
\tag{S12}
\label{eq:S12}
\end{equation}

In the chasing-pair model, there is no interspecific information use(\(l_2=0\)), and then the Eq.~\ref{eq:S12} can be rewritten as:
\begin{equation}
f_i(R^\text{(F)}) = \frac{w_ik_iR^\text{(F)}}{R^\text{(F)} + (d_i + k_i)/a_i},\  (i=1,2),
\tag{S13}
\label{eq:S13}
\end{equation}
where, according to Eq. 2, \(a_1=a_1'\). Combining the first and second sub-equations of Eq.~1, we can obtain the expression reported in the literature~\cite{WangXSI2020}:
\begin{equation}
\left\{
\begin{aligned}
\dot{C}&_1 = (f_1(R^\text{(F)}) - D_1) C_1,\\
\dot{C}&_2 = (f_2(R^\text{(F)}) - D_2) C_2.
\end{aligned}
\right.
\tag{S14}
\label{eq:S14}
\end{equation}

Combining Eq.~1 and Eq.~\ref{eq:S12}, Eq.~\ref{eq:S14} can be rewritten as the information-use equation:
\begin{equation}
\left\{
\begin{aligned}
\dot{C}&_1 = (F_1(R^\text{(F)},C_2) - D_1) C_1,\\[6pt]
\dot{C}&_2 = (F_2(R^\text{(F)}) - D_2) C_2.
\end{aligned}
\right.
\tag{S15}
\label{eq:S15}
\end{equation}

At steady state ($\dot{C}_1 = 0$, $\dot{C}_2 = 0$), by examining Eq.~\ref{eq:S14}, we can see that a single unknown variable corresponds to two equations, making the system generally unsolvable. However, the first sub-equation in Eq.~\ref{eq:S15} introduces an additional variable $C_2$, so that the two equations now correspond exactly to two unknowns, rendering the system solvable. 
This result demonstrates that interspecific information use enables coexistence among species.

\section{Dimensional analysis for the scenario involving information aid}
\label{Dimensional analysis}
Combining Eqs. 1 and 2, we obtain the population dynamics for biotic and abiotic resource systems that incorporate information use:

\begin{equation}
\left\{
\begin{aligned}
\dot{x}_1 &= a_1' \left[ 1 + l_2 \frac{C_2}{K_2 + C_2} \right] C_1^\text{(F)} R^\text{(F)} - (k_1 + d_1)x_1, \\
\dot{x}_2 &= a_2 C_2^\text{(F)} R^\text{(F)} - (k_2 + d_2)x_2, \\
\dot{C}_1 &= w_1 k_1 x_1 - D_1 C_1, \\
\dot{C}_2 &= w_2 k_2 x_2 - D_2 C_2, \\
\dot{R}   &= g(\{R\},\{x_i\},\{C_i\})\ (i=1,2).
\end{aligned}
\right.
\label{eq:S16}
\tag{S16}
\end{equation}

For biotic resources, \( g(\{R\}, \{x_i\}, \{C_i\}) = \eta R(1 - R/K_R) - k_1 x_1 - k_2 x_2 \), whereas for abiotic resources, 
and \( g(\{R\}, \{x_i\}, \{C_i\}) = \zeta (1 - R/K_R) - k_1 x_1 - k_2 x_2 \). 
In this framework, \( C_i = C_i^\text{(F)} + x_i \),  and \( R = R^\text{(F)} + x_1 + x_2 \) denote the total abundances of consumers and resources, respectively.
Eq.~\ref{eq:S16} already contains several dimensionless quantities, including \( x_1, x_2, C_1^\text{(F)}, C_2^\text{(F)}, R^\text{(F)}, C_1, C_2, R, l_2, K_2, w_1, \) and \( w_2 \). 
To ensure full nondimensionalization, we introduce a scaled time variable \( \tilde{t} = t / \tau \), where \( \tau = \tilde{D}_1 / D_1 \) and \( \tilde{D}_1 \) is a reducible dimensionless constant that can take any positive value. 
We then define the corresponding dimensionless parameters as \( \tilde{a}_1 = a_1 \tau \), \( \tilde{a}_2 = a_2 \tau \), \( \tilde{k}_1 = k_1 \tau \), \( \tilde{k}_2 = k_2 \tau \), \( \tilde{d}_1 = d_1 \tau \), \( \tilde{d}_2 = d_2 \tau \), \( \tilde{D}_1 = D_1 \tau \), and \( \tilde{D}_2 = D_2 \tau \). 
Additionally, we set \( \tilde{\zeta} = \xi \) and \( \tilde{\eta} = \eta \). 
Substituting these nondimensional variables and parameters into Eq.~\ref{eq:S16} yields the following expression:

\begin{equation}
\left\{
\begin{aligned}
\dot{x}_1 &= \tilde{a}_1' \left[ 1 + l_2 \frac{C_2}{K_2 + C_2} \right] C_1^\text{(F)} R^\text{(F)} - (\tilde{k}_1 + \tilde{d}_1)x_1, \\
\dot{x}_2 &= \tilde{a}_2 C_2^\text{(F)} R^\text{(F)} - (\tilde{k}_2 + \tilde{d}_2)x_2, \\
\dot{C}_1 &= w_1 \tilde{k}_1 x_1 - \tilde{D}_1 C_1, \\
\dot{C}_2 &= w_2 \tilde{k}_2 x_2 - \tilde{D}_2 C_2, \\
\dot{R}   &= g(\{R\},\{x_i\},\{C_i\})\ (i=1,2).
\end{aligned}
\right.
\label{eq:S17}
\tag{S17}
\end{equation}

Unless noted otherwise, the tilde symbol “$\sim$” is omitted, and all variables and parameters are considered dimensionless in the simulations.
 
\section{Simulation Procedure of the Stochastic Simulation Algorithm}
\label{Simulation Procedure}
To quantitatively assess the effect of stochasticity, we employed the stochastic simulation algorithm (SSA)~\cite{GillespieSI2007}, with all stochastic simulations performed according to the standard Gillespie procedure~\cite{GillespieSI2007}.

\section{Simulation details of the main text figures}
\label{Simulation details}
In Figs. 1(C) and (E), $a_1 = 0.2$, $a_2 = 0.2$, $l_2 = 0$, $w_1 = 0.018$, $w_2 = 0.018$, $D_1 = 0.017$, $D_2 = 0.01$, $K_2 = 40$, $\zeta = 55$, $K_R = 55$, $k_1 = 4.5$, $k_2 = 4.5$, $d_1 = 150$, and $d_2 = 150$. 
In Figs. 1(D) and (F), all parameters are identical to those in (C) and (E), except that $l_2 = 2$.

In Figs. 2A, \( a_1 = 0.2 \), \( a_2 = 0.2 \), \( l_2 = 0.5 \), \( K_2 = 40 \), 
\( w_1 = 0.02 \), \( w_2 = 0.02 \), \( D_1 = 0.011 \), \( D_2 = 0.010 \), 
\( \zeta = 55 \), \( K_R = 10000 \), \( k_1 = 4.5 \), \( k_2 = 4.5 \), 
\( d_1 = 1 \), and \( d_2 = 1 \). In Fig. 2B, \( a_1 = 0.01 \), \( a_2 = 0.01 \), \( l_2 = 0.5 \), \( K_2 = 200 \), 
\( w_1 = 0.15 \), \( w_2 = 0.15 \), \( D_1 = 0.024 \), \( D_2 = 0.020 \), 
\( \eta = 0.03 \), \( K_R = 10000 \), \( k_1 = 0.8 \), \( k_2 = 0.8 \), 
\( d_1 = 80 \), and \( d_2 = 80 \). In Fig. 2C, \( a_1 = 0.01 \), \( a_2 = 0.01 \), \( l_2 = 0.5 \), \( K_2 = 200 \), 
\( w_1 = 0.15 \), \( w_2 = 0.15 \), \( D_1 = 0.023 \), \( D_2 = 0.020 \), 
\( \eta = 0.1 \), \( K_R = 3000 \), \( k_1 = 0.8 \), \( k_2 = 0.8 \), 
\( d_1 = 80 \), and \( d_2 = 80 \). In Fig. 2D, \( a_1 = 0.01 \), \( a_2 = 0.01 \), \( l_2 = 0.5 \), \( K_2 = 200 \), 
\( w_1 = 0.15 \), \( w_2 = 0.15 \), \( D_1 = 0.021 \), \( D_2 = 0.020 \), 
\( \eta = 0.01 \), \( K_R = 10000 \), \( k_1 = 0.8 \), \( k_2 = 0.8 \), 
\( d_1 = 80 \), and \( d_2 = 80 \). In Fig. 2E, \( a_1 = 0.01 \), \( a_2 = 0.01 \), \( l_2 = 0.5 \), \( K_2 = 200 \), 
\( w_1 = 0.15 \), \( w_2 = 0.15 \), \( D_1 = 0.021 \), \( D_2 = 0.020 \), 
\( k_1 = 0.8 \), and \( k_2 = 0.8 \). In Fig. 2F, \( a_1 = 0.2 \), \( a_2 = 0.2 \), \( l_2 = 2 \), \( K_2 = 40 \), 
\( w_1 = 0.02 \), \( w_2 = 0.02 \), \( D_2 = 0.010 \), \( k_1 = 4.5 \), 
\( k_2 = 4.5 \), \( d_1 = 60 \), and \( d_2 = 60 \).
Figs. 2A– 2D were generated through ODE simulations and fixed-point calculations based on Eqs. 1- 3. For Figs. 2E– 2F, the stability of the fixed points was examined using the Jacobian matrix, as described in the SM~sec. \ref{Stability analysis}.

In Figs. 3A and 3C, \( a_1 = 0.005~\mathrm{h^{-1}} \), \( a_2 = 0.006~\mathrm{h^{-1}} \), 
\( l_2 = 0.5 \), \( K_2 = 650 \), \( w_1 = 0.15 \), \( w_2 = 0.15 \), 
\( D_1 = 0.03~\mathrm{h^{-1}} \), \( D_2 = 0.03~\mathrm{h^{-1}} \), 
\( \eta = 0.32~\mathrm{h^{-1}} \), \( K_R = 7000 \), 
\( k_1 = 0.8~\mathrm{h^{-1}} \), \( k_2 = 0.8~\mathrm{h^{-1}} \), 
\( d_1 = 60~\mathrm{h^{-1}} \), and \( d_2 = 60~\mathrm{h^{-1}} \).  
Since the investigated subjects are cliff-nesting seabirds, 
and exhibit alternating foraging and chick-rearing behavior, as well as a prey preference for surface-schooling fish, 
the parameter settings may deviate slightly from field conditions. In Figs. 3B and 3D, \( a_1 = 0.2~\mathrm{day^{-1}} \), \( a_2 = 0.2~\mathrm{day^{-1}} \), 
\( l_2 = 2 \), \( K_2 = 40 \), \( w_1 = 0.02 \), \( w_2 = 0.02 \), 
\( D_1 = 0.017~\mathrm{day^{-1}} \), \( D_2 = 0.01~\mathrm{day^{-1}} \), 
\( \zeta = 55~\mathrm{day^{-1}} \), \( K_R = 10000 \), 
\( k_1 = 4.5~\mathrm{day^{-1}} \), \( k_2 = 4.5~\mathrm{day^{-1}} \), 
\( d_1 = 150~\mathrm{day^{-1}} \), and \( d_2 = 150~\mathrm{day^{-1}} \). In this case, consumer species \( C_1 \) and \( C_2 \) correspond to \textit{T.~castaneum} and \textit{T.~confusum}, respectively. In Figs. 3E, \( a_1 = 0.3~\mathrm{day^{-1}} \), \( a_2 = 0.3~\mathrm{day^{-1}} \), 
\( l_2 = 2 \), \( K_2 = 100 \), \( w_1 = 0.018 \), \( w_2 = 0.018 \), 
\( K_R = 20000 \), \( k_1 = 5~\mathrm{day^{-1}} \), \( k_2 = 5~\mathrm{day^{-1}} \), 
\( d_1 = 60~\mathrm{day^{-1}} \), and \( d_2 = 60~\mathrm{day^{-1}} \).  
Notably, for GRP1, \( \zeta = 60~\mathrm{day^{-1}} \), \( D_1 = 0.028~\mathrm{day^{-1}} \), \( D_2 = 0.023~\mathrm{day^{-1}} \); for GRP2, \( \zeta = 65~\mathrm{day^{-1}} \), \( D_1 = 0.025~\mathrm{day^{-1}} \), \( D_2 = 0.02~\mathrm{day^{-1}} \).

In Figs. S1A and C, $a_1 = 0.0007$, $a_2 = 0.0007$, $l_2 = 0$, $w_1 = 0.3$, $w_2 = 0.3$, $D_1 = 0.021$, $D_2 = 0.02$, $K_2 = 200$, $\eta = 0.8$, $K_R = 400$, $k_1 = 0.8$, $k_2 = 0.8$, $d_1 = 2$, and $d_2 = 2$. In Figs. S1B and D, all parameters are identical to those in Figs. S1C and E, except that $l_2 = 0.5$.

\newpage
\renewcommand{\thetable}{S\arabic{table}}
\setcounter{table}{0}
\begin{table}[htbp]
\centering
\caption{List of symbols and their definitions.}
\begin{tabularx}{0.95\textwidth}{lX}
\hline
\textbf{Symbols} & \textbf{Illustrations / Definitions} \\
\hline
\( C_i \) & The population abundance of consumer species \( C_i \). \\
\( R \) & The population abundance of resource species \( R \). \\
\( C_i^\text{(F)} \) & The freely wandering individuals of consumer species \( C_i \). \\
\( R^\text{(F)} \) & The freely wandering individuals of resource species \( R \). \\
\( C_i^{\text{(A)}} \) & The steady-state analytical solution for consumer abundances \( C_i^{\text{(A)}} \). \\
\( R^{\text{(A)}} \) & The steady-state analytical solution for resource abundances \( R^{\text{(A)}} \). \\
\( x_i \) & The chasing pair \( C_i^{\text{(P)}} \lor R^{\text{(P)}} \) formed by a consumer and a resource. \\
\( a_1' \) & The encounter rate of the predation process in the absence of social information transfer. \\
\( a_i \) & The actual encounter rate of the predation process with information aid. \\
\( l_2 \) & The maximum relative increase in searching efficiency. \\
\( K_2 \) & The half saturation constant at which this increase levels off. \\
\( k_i \) & The capture rate within a chasing pair. \\
\( d_i \) & The escape rate within a chasing pair. \\
\( w_i \) & The biomass conversion ratio from species \( R \) to \( C_i \). \\
\( D_i \) & The mortality rate of consumer species \( C_i \). \\
\( \eta \) & The intrinsic growth rate. \\
\( \zeta \) & The external supply rate of resource \( R \). \\
\( \kappa_0 \) & The carrying capacity. \\
\( \kappa_R \) & The equilibrium abundance maintained in the absence of consumers. \\
\hline
\end{tabularx}
\label{tab:symbols}
\end{table}

\clearpage
\section{Supplemental Figures}
\begin{figure}[h]
	\centering
	\includegraphics[width=12cm]{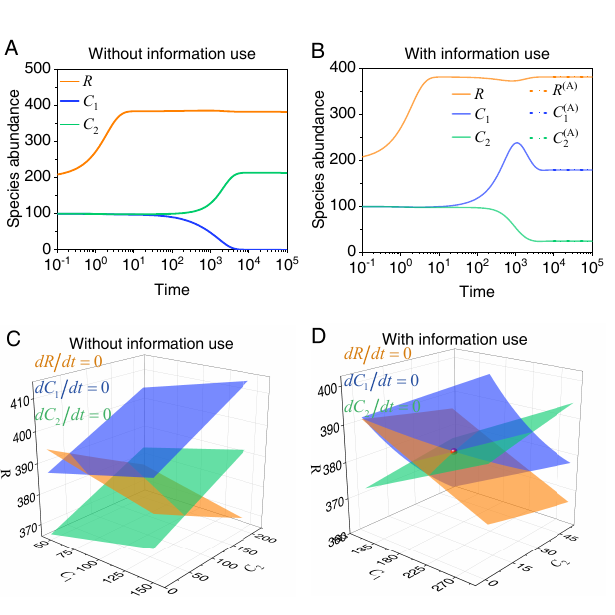}
	\caption{\label{stability} ODE results for two consumer species competing for a single biotic resource type. (A-B) Time courses of two consumer species competing for a single biotic resource species. (C-D)Positive solutions to the steady-state equations (see Eqs.1-3): $\dot R = 0$ (orange surface), $\dot C_1 = 0$ (blue surface), and $\dot C_2 = 0$ (green surface), representing the zero-growth isoclines. The red dot represents the stable fixed point. The dotted curves in (D) correspond to the analytical steady-state abundances, labeled with the superscript '(A)'. See SM Sec. VI and Table S1 for simulation details of Figs. S1–2.}	
\end{figure}

\begin{figure}[h]
	\centering
	\includegraphics[width=12cm]{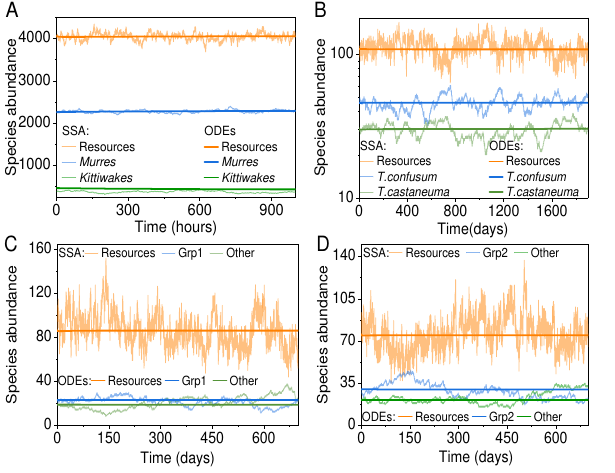}
	\caption{\label{SI2} The SSA and ODE simulations results including resource dynamics.(A) Coexistence dynamics between Kittiwakes and Murres within a seabird community under a single biotic resource.(B) Coexistence dynamics of flour beetles (T. confusum and T. castaneum) under a single abiotic resource. (C–D) Abundance variations of two Drosophila serrata groups (Grp1 and Grp2) and other minor taxa when sharing a single abiotic resource. The parameter settings and simulation methods: (A) and Fig.3C are identical; (B) and Fig.3D are identical; The relevant settings for (C) and (D) are described in Fig.3E.}	
\end{figure}

\end{document}